 \documentclass[10pt,preprint]{aastex}  
 \usepackage{natbib}
\def\ang{\AA}

\def\gapprox{\lower.4ex\hbox{$\;\buildrel >\over{\scriptstyle\sim}\;$}}
\def\lapprox{\lower.4ex\hbox{$\;\buildrel <\over{\scriptstyle\sim}\;$}}

\shortauthors{ASCHWANDEN}
\shorttitle{Self-Organized Criticality Systems}

\begin{document}

\title{ Non-Stationary Fast-Driven Self-Organized Criticality 
	in Solar Flares}

\author{Markus J. Aschwanden}

\affil{	Solar and Astrophysics Laboratory,
	Lockheed Martin Advanced Technology Center, 
        Dept. ADBS, Bldg.252, 3251 Hanover St., Palo Alto, CA 94304, USA; 
        (e-mail: \url{aschwanden@lmsal.com})}

\begin{abstract}
The original concept of self-organized criticality 
(Bak et al.~1987), applied to solar flare statistics 
(Lu and Hamilton 1991), assumed a slow-driven and 
stationary flaring rate, which warrants time scale
separation (between flare durations and inter-flare
waiting times), it reproduces power-law distributions for 
flare peak fluxes and durations, but predicts an 
exponential waiting 
time distribution. In contrast to these classical
assumptions we observe: (i) multiple energy dissipation 
episodes during most flares, (ii) violation of the 
principle of time scale separation, (iii) a fast-driven 
and non-stationary flaring rate, (iv) a power law 
distribution for waiting times $\Delta t$, with a slope 
of $\alpha_{\Delta t} \approx 2.0$, as predicted from 
the universal reciprocality between mean flaring rates 
and mean waiting times; and (v) pulses with rise times
and decay times of the dissipated magnetic free energy on time
scales of $12\pm6$ min, up to 13 times in long-duration
($\lapprox 4$ hrs) flares. These results are inconsistent  
with coronal long-term energy storage (Rosner and Vaiana 1978),
but require photospheric-chromospheric current injections 
into the corona.
\end{abstract}


\section{		INTRODUCTION 			}

{\sl Self-organized criticality (SOC)} models are extremely useful 
to obtain physical scaling laws from the statistics of nonlinear
energy dissipation processes (for reviews see 
Charbonneau et al.~2001; 
Aschwanden 2011a, 2019; 
Pruessner 2012; 
Charbonneau 2013;
Aschwanden et al.~2016a, and references therein).
The original concept of avalanches that occur randomly above some
threshold, triggered by continuously dripped sand grains on top
of a sandpile in a critical state, is due to Bak et al.~(1987), 
while the first applications to solar flare statistics have been 
explored by Lu and Hamilton (1991) and Crosby et al.~(1993).
SOC models can be tested by the power law distributions of various
geometric, temporal, and other physical parameters, which should
reveal power law slopes that are consistent with the underlying
physical scaling laws, as well as with the functional shape of their
waiting time distributions. In a slow-driven SOC model, avalanches
occur rarely and are temporally separated, a condition that is
called {\sl time scale separation}, while this condition is likely
to be violated in fast-driven SOC systems. In this study we will
demonstrate that the time scale separation is clearly infringed
in the case of solar flares. While the energy build-up or storage
time is much longer than the duration of an avalanche in a classical
slow-driven SOC system (Fig.~1a), we find here that the energy 
build-up or storage time is comparable with the energy dissipation
time (of free magentic energy), being a fraction of the flare 
duration only (Fig.~1b), and thus be much shorter than the waiting
time between two subsequent flares. 

This new aspect of fast-driven SOC systems has some far-reaching 
consequences that have virtually not been investigated yet. 
The power law slope of flare durations, which typically has a value
of $\alpha_T \approx 2.0$, revealed higher values of $\alpha_T
\approx 2.0-5.0$ during solar cycle maximum years, which was
interpreted as a solar-cycle dependent flare pile-up bias
(Aschwanden 2011a, 2011b, 2011c; Aschwanden and Freeland 2012). 
Sufficiently fast-driven sandpile models produce frequent 
occurrencies, where many avalanches mutually overlap in time 
and identification and definition of single events becomes
problematic due to the violation of time scale separation. 
This confusion problem is particularly problematic for 
1-dimensional data (such as light curves of flares at a given
wavelength), but is much alleviated in 2-dimensional data
(where near-simultaneous events can be separated in space
and time). 

Another testbed of fast-driven SOC models is the waiting time
(or inter-flare time interval) distribution. In classical 
slow-driven sandpile SOC models, individual avalanches occur
independently of each other, which predicts an exponential
(Poissonian) waiting time distribution for stationary flaring rates. 
However, non-stationary flaring rates produce different
waiting time distributions, which depend on the variability
function of the flaring rate 
(Wheatland et al.~1998; Wheatland 2000c, 2002, 2006). 
In this study we
emphasize the novel model of the Poissonian non-stationary waiting
time distribution that is based on the universal reciprocality
of flaring rates and mean flare waiting times (Aschwanden
and McTiernan 2010). This universal model has no free parameters 
(except for a normalization constant) and predicts a power law slope of 
$\alpha_{\Delta t}=2$, which mostly agrees with the observations
of solar flares (Wheatland et al.~1998; Wheatland 2003; Moon et al.~2001; 
Kanazir and Wheatland 2010; Aschwanden and McTiernan 2010).

The goal of this study is a deeper understanding of non-standard
SOC models, addressing slow-driven versus fast-driven SOC models,
the time scale separation, the multiplicity of energy release 
pulses during a single avalanche event, the violation of time scale 
separation, and the universal Poissonian non-stationary waiting 
time distribution.
We present a brief description of the analytical theory of waiting
time distributions (Section 2), observations and data analysis of
solar flare data (Section 3), a discussion of the new findings
(Section 4), and conclusions (Section 5).

\section{		THEORY 		}

\subsection{		Analytical Waiting Time Distribution	}

Waiting times $\Delta t$, the inter-event time intervals between
two subsequent events of a Poissonian point process are expected 
to exhibit an exponential function in the case of a stationary 
random process. The time series sample may consist of 
time intervals observed in statistically independent events and
sampled at different locations and times.
Thus the probability distribution function
$p(\Delta t)$ is defined by, 
\begin{equation}
	p(\Delta t) = \lambda_0\ \exp^{-\lambda_0 \Delta t} \ ,
\end{equation} 
where $\lambda_0$ represents the mean event occurrence rate,
and the distribution is normalized to unity, i.e., 
$\int_0^\infty p(\Delta t) d\Delta t=1$. A random process
can be called a {\sl stationary Poisson process} when the
average flaring rate $\lambda_0$ is time-independent and stays
constant as a function of time.

A more general approach of waiting time distributions is the concept
of inhomogeneous or {\sl non-stationary Poisson processes}, where
the mean flaring rate $\lambda(t)$ becomes a function of time itself
(e.g., Jaynes 2003; Sivia and Skilling 2006; Scargle 1998; 
Wheatland et al.~1998, 2000; Litvinenko and Wheatland 2001;
Wheatland and Litvinenko 2002). 
Applying {\sl Bayesian statistics}, a time series can be subdivided
into {\sl Bayesian blocks}, during which the occurrence rate $\lambda_i$ 
is assumed to be piece-wise stationary during a time interval 
$[t_i, t_{i+1}]$,
\begin{equation}
	p(t, \Delta t) = \left\{ \begin{array}{cc}
		\lambda_1\ \exp^{-\lambda_1 \Delta t} & {\rm for}\ t_1<t<t_2 \\ 
		\lambda_2\ \exp^{-\lambda_2 \Delta t} & {\rm for}\ t_2<t<t_3 \\ 
		.......                              &                      \\
		\lambda_n \exp^{-\lambda_n \Delta t} & {\rm for}\ t_n<t<t_{n+1}  
		\end{array} \right. \ ,
\end{equation}	
The summation of the piece-wise Bayesian blocks over discrete time intervals
can be converted into a continuous integral function, 
\begin{equation}
	p(\Delta t) = \int_0^T \lambda(t)\ p(t,\Delta t) \ dt \ ,
\end{equation}
where the probability $p(t,\Delta t)$ in each Bayesian block is weighted
by the number of events $\lambda(t)$. The total duration of the time series
is $T$, and the normalization is given by the total number of events,
i.e., $N = \int_0^T \lambda(t) \ dt$. 
Inserting the time-dependent probability $p(t,\Delta t) = \lambda(t) \
\exp^{-\lambda(t) \Delta t)}$ into Eq.~(3) yields,
\begin{equation}
	p(\Delta t) = {\int_0^T \lambda(t)^2\ \exp{^{-\lambda(t) \Delta t}}
			dt \over
		       \int_0^T \lambda(t) dt } \ .
\end{equation}
Following Wheatland et al.~(1998, 2000), we substitute the time variable $t$
with the event occurrence rate $\lambda$, by defining the function
$f(\lambda)=(1/T) dt(\lambda)/d\lambda$, which is equivalent to
$f(\lambda) d\lambda = dt/T$, 
\begin{equation}
	p(\Delta t) = {\int_0^\infty 
	f(\lambda)\ \lambda^2\ \exp{^{-\lambda \Delta t}} d\lambda \over
		       \int_0^\infty 
	\lambda f(\lambda) d\lambda } \ .
\end{equation}

We make now a special choice for the flaring rate distribution $f(\lambda)$
that contains (i) a reciprocal relationship $f(\lambda) \propto \lambda^{-1}$
for small flaring rates $\lambda \lapprox \lambda_0$, and (ii) contains an
exponential drop-off at large flaring rates $\lambda \gapprox \lambda_0$
(see also Eq.~5.2.16 in Aschwanden 2011a),
\begin{equation}
	f(\lambda) = 
	\lambda^{-1} \ \exp{ \left( - { \lambda \over \lambda_0} \right)} .
\end{equation}
The scale-free range of $\lambda < \lambda_0$ is visualized in Fig.~2 
(left panel, solid line), together with the exponential component 
(Fig.~2, left panel, dashed line).
The scale-free property with the scaling $f(\lambda) \propto \lambda^{-1}$
is easy to understand, because the number of
events $f(\lambda)$ is proportional to the mean waiting time $<\Delta t>$, 
which in turn is reciprocal to the mean flaring rate $<\lambda>$, 
e.g., $f(\lambda) \propto <\lambda^{-1}> \propto <\Delta t>$, and
thus is universally valid for every waiting time distribution. 
In addition, the exponential term
in Eq.~(6) essentially produces an upper boundary of the reciprocal
function at $\lambda \gapprox \lambda_0$. The expression given in Eq.~6
fulfills also the normalization $\int_0^\infty \lambda \ f(\lambda)
\ d\lambda = \lambda_0$. The waiting time distribution (Eq.~5) can 
then be written as,
\begin{equation}
	p(\Delta t) = \int_0^\infty
	\left( {\lambda \over \lambda_0} \right)
	\exp{ \left( - {\lambda \over \lambda_0}
	[1 + \lambda_0 \Delta t]  \right) } \ d\lambda ,
\end{equation}
which, with defining $a=-(1+\lambda_0 \Delta t)/\lambda_0$,
corresponds to the integral $\int xe^{ax} dx
=(e^{ax}/a^2)(ax-1)$ and becomes $\int_0^\infty x e^{ax} dx = 1/a^2$
when integrated over $[0<x<\infty]$, yielding the solution 
$p(\Delta t)=1/(a^2 \lambda_0^2)$, and we obtain for the 
waiting time distribution,
\begin{equation}
	p(\Delta t) = {\lambda_0 \over (1 + \lambda_0 \Delta t)^2} \ .
\end{equation}
Note that this waiting time distribution contains no free variables,
except for the normalization constant $\lambda_0$. Thus, this model
predicts universally a power law slope of $\alpha_{\Delta t}=2$ for 
any waiting time distribution. The only unterlying assumption is
the reciprocality of flaring rates and waiting times, which naturally
emerges from the property of scale-freeness in self-organized
criticality models (Aschwanden and McTiernan 2010).

A comparison of stationary and non-stationary waiting time distributions
is shown in Fig.~3, as well as for a slow-driven and fast-driven
SOC model (Fig.~3). Note the reciprocal relationship between the
flare occurrence rate (y-axis in Fig.~3) and the waiting time
(x-axis in Fig.~3), differing by a factor of $10^2$.  
A parametric set of theoretically predicted waiting times with
various values of $\lambda_0=0.02, ..., 0.12$ is shown in Fig.~4c.

\subsection{	Numerical Simulations of Waiting Time Distributions	}

It is customary to perform Monte-Carlo simulations of waiting time
distributions $N(\Delta t) d\Delta t$ or occurrence frequency distributions
$N(x) dx$ by random generator values $x=x_1, x_2, ..., x_n$ that have
a prescribed function of their frequency distribution. Examples for
exponential and power-law distributions are given in Section 7.1.4
of Aschwanden (2011a). The normalization is given by the integral
of the probability function $p(x)$,
\begin{equation}
	\int_0^\infty p(x)\ dx = 1 \ .
\end{equation}
The total probability $\rho(x)$ to have a value in the range of
$[0, x]$ is then the integral,
\begin{equation}
	\rho(x) = \int_0^x p(x')\ dx' \ .
\end{equation}
Then we invert the integral function $\rho(x)$ and denote it with
the analytical inverse function $\rho^{-1}$, so that
\begin{equation}
	x = \rho^{-1}(\rho) = \rho^{-1}(\rho[x]) \ ,
\end{equation}
yields a transform that allows us to generate values $x_i$
from a distribution of probability values  $\rho_i$. There are
many numerical random generator algorithms available that produce
a random number $\rho_i$ in a homogeneous range of $[0,1]$,
which can then be used to generate values $x_i$ with the
mapping transform $x_i = \rho^{-1}(\rho_i)$. The frequency
distribution of these values $x_i$ will then fulfill the
prescribed function $p(x)$.  

In our case we want to simulate the waiting time distribution
function that is given by the probability function $p(\Delta t)$
(Eq.~8),
\begin{equation}
	p(\Delta t) = {\lambda_0 \over (1 + \lambda_0 \Delta t)^2} \ ,
\end{equation}
which fulfills the normalization
\begin{equation}
	\int_0^1 p(\Delta t) \ d\Delta t = 1 \ . 
\end{equation}
The integral function $\rho(\Delta t)$ of the probability function
$p(\Delta t)$ is then 
\begin{equation}
	\rho(\Delta t) = 
	\int_0^{\Delta t} {\lambda_0 \over (1 + \lambda_0 \Delta t')^2} 
	d \Delta t' = {\lambda_0 \Delta t \over 1 + \lambda_0 \Delta t} \ ,
\end{equation}
The inversion of the probability function $\rho(\Delta t)$ is then simply
\begin{equation}
	\Delta t = {\rho \over \lambda_0 (1 - \rho)}
\end{equation}
which can be used to simulate a set of waiting times $\Delta t_i$
using random numbers $\rho_i$ in the homogeneous range $[0,1]$,
\begin{equation}
	\Delta t_i = {\rho_i \over \lambda_0 (1 - \rho_i)} \ ,
	\quad {\rm for}\ [0 < \rho_i < 1] \ .
\end{equation}
Such a simulation for $N=575$ events and $\lambda_0=1.7$ is shown 
in Fig.~4 (middle panel), along with the theoretical distribution
function $p(\Delta t)$ (Eq.~8).
 
\section{		DATA ANALYSIS			}

\subsection{	Observations and Data Selection 	}

We analyzed the same data set of 170 solar flares presented in 
Aschwanden et al.~(2014a), which includes all M- and X-class flares 
observed with the {\sl Solar Dynamics Observatory (SDO)} 
(Pesnell et al.~2011) during the first 
3.5 years of the mission This selection of events has a
heliographic longitude range of $[-45^\circ, +45^\circ]$, for which
magnetic field modeling can be faciliated without too severe
foreshortening effects near the solar limb. We use the 45-s
line-of-sight magnetograms from the 
{\sl Helioseismic and Magnetic Imager (HMI)}/SDO and make use of all
coronal extreme-ultraviolet (EUV) channels of the
{\sl Atmospheric Imaging Assembly (AIA)}/SDO (in the six wavelengths
94, 131, 171, 193, 211, 335 \ang ), which are sensitive to
strong iron lines (Fe VIII, IX, XII, XIV, XVI, XVIII, XXI, XXIV)
in the temperature range of $T \approx 0.6-16$ MK. 
For most of the analysis we analyzed images with a cadence of 6 min,
but present one event with the full AIA time cadence of 12 s. 

\subsection{	Magnetic Field Computations		}

The coronal magnetic field is modeled by using the line-of-sight
magnetogram $B_z(x,y)$ from HMI/SDO and (automatically detected)
projected loop coordinates $[x(s), y(s)]$ in each EUV wavelength of AIA.
A full 3-D magnetic field model ${\bf B}(x,y,z)$ is computed
for each time interval and flare with a cadence of 6 min (0.1 hrs). 
The total duration of a flare is defined by the GOES flare start 
and end times, including a margin of 0.5 hrs before and after 
each flare. The magnetic field is computed 
with the {\sl vertical-current approximation non-linear force-free
field (VCA-NLFFF)} code, which is described for the original 
first version (Aschwanden 2013), and has been improved in accuracy
in the second (Aschwanden et al.~2016b) and third (VCA3-NLFFF) 
version (Aschwanden 2020). 

\subsection{	Time Evolution of Free Energy		}

The main physical parameter that we are interested in here 
is the time evolution of the free energy, which is defined 
as the difference between the potential and non-potential 
magnetic field, i.e., $E_{free}(t)=E_{np}(t)-E_p(t)$.

We show the time evolution of the free energy $E_{free}(t)$
for 20 flare events (out of the 170 analyzed events) in Figs.~5 to 7.
We decompose the time profiles into pulses that consist of
a rise time phase $\tau_{rise}=t_p-t_s$, and a decay time
phase $\tau_{decay}=t_e-t_p$. The peak times $t_p$
are measured at the local maxima of the time evolution function
$E_{free}(t)$, and the starting times $t_s$ and end times are
derived from the local minima preceding and following 
each peak time. For clarity we represent the decay phases of
the pulses with grey areas in Figs.~5 and 6. 
In Fig.~5 we show relatively 
simple flare events with one single ($n_p=1$) or two peaks 
($n_p=2$), while the 10 cases shown in Fig.~6 were selected
from the flare events with the longest duration, which
exhibit from $n_p=5$ to $n_p=13$ peaks.

In Fig.~7 we show the time profiles of the free energy $E_{free}(t)$
with higher time resolution: The nominal resolution is 6 min
(Fig.~7b), an intermediate resolution is 1 min (Fig.~7c), and
the full time resolution of AIA is 12 s (Fig.~7d). 
The fluctuations visible
at the highest cadence (Fig.~7d) show a mean and standard deviation
of $E_{free}=(42 \pm 7) \times 10^{30}$ erg, which indicates 
uncertainties of $\sigma_E \approx 7/42 \approx 0.17$. This
uncertainty in the free energy includes numerical noise, mostly
caused by the decomposition of unipolar magnetic charges from
the HMI magnetograms and from the automated detection of coronal
loops in the AIA images. Nevertheless, the time profiles shown in
Fig.~7 reveal about 1-3 significant pulses for this event, 
while Fig.~6 shows
5-13 significant energy dissipation pulses per flare. 

\subsection{	Statistics of Time Scales		}

Statistics of time scales is given in Fig.~8. The number of
energy dissipation pulses per flare ranges from $n_p=1$ to
$n_p=13$ (see Figs.~5 and 6), as derived from the (slightly smoothed) 
time profiles of the free energy, $E_{free}(t)$. Each of the pulses
is characterized by the rise time (which can be interpreted as
magnetic energy loading time by new flux emergence), 
$\tau_{rise}=0.1-1.2$ hrs = 6-72 min (Fig.~8a), 
by the pulse decay time (which can be interpreted as magnetic energy 
dissipation time), $\tau_{decay}=0.1-0.7$ hrs = 6-42 min (Fig.~8b), 
and the total pulse duration $\tau_{pulse}=0.2-1.5$ hrs = 12-90 min 
(Fig.~8c).
The lower limit of $\tau_{rise,min}=\tau_{decay,min}=0.1$ hrs = 6 min 
is caused by the chosen cadence in the calculation of magnetic energies.

The flare duration times have a range of $\tau_{flare}=1.1-5.2$ hrs
(Fig.~8d), which is about an order of magnitude longer than the 
pulse rise or decay times. This difference can be explained by
the fact that the time scale of magnetic energy dissipation, which
is similar to the duration of hard X-ray emission, is generally
shorter than the time scale of soft X-ray emission, that was used
by NOAA to define the flare duration.

Finally, we also measure the waiting times of flare events,
using all GOES M- and X-class flare events during the first 3.5 years
of the SDO mission (from 2010 June 12 to 2014 Nov 16), including
those events near the limb for which magnetic modeling was not feasible.
The range of waiting times derived from the starting time difference
of these 575 flares covers $\Delta t=0.2-2000$ hrs (Fig.~8e). 
Note that truncation effects due to the solar rotation and the 
selected longitudinal range ($\pm 45^\circ$) are ignored in the 
waiting time statistics here, although it could affect the correct 
waiting time measurement for events near the east or west limb.
The waiting 
time distribution forms a power law distribution with a slope of
$\alpha_{\Delta t}=2.0$ for time scales of $\Delta t \gapprox 1$ hr
(Fig.~8e) and closely follows the predicted function derived
theoretically (Eq.~8) for a normalization constant of 
$\lambda_0=0.07$ hr$^{-1}$. According to the definitions of waiting
times $\Delta t$, energy storage times $\tau_{storage} \approx \tau_{rise}$, 
and energy dissipation times $\tau_{diss} \approx \tau_{decay}$
given in Fig.~1, most of the storage times (Fig.~8a) are much
shorter than the waiting times (Fig.~8e), and thus are consistent
with the fast-driven SOC model (Fig.~1b), rather than with the
slow-driven SOC model (Fig.~1a).

\subsection{	Correlation of Free Energy with Hard X-rays 	}

If the magnetic free energy is the main energy input in
solar flares, and the energy converted into acceleration of (nonthermal)
particles $E_{nth}$ conveys the major energy output, we would expect some 
correlation between the free energy time profile $E_{free}(t)$ 
and the hard X-ray flux time profile $F_{HXR}(t)$, which most easily 
can be inferred from the time derivative of the GOES soft X-ray
time profile, i.e., $F_{HXR} = \partial F_{HXR}(t)/\partial t$,
according to the Neupert effect (Neupert 1968; Dennis and Zarro 1993).

We juxtapose these two time profiles $E_{diss}(t)$ and $F_{HXR}(t)$
for 20 flare events in Figs.~5 and 6, where the time profiles of
the energy dissipation (inferred from the pulse decay time intervals
marked with grey areas) and the GOES 1-8 \ang\ flux (marked with
hatched areas) are shown. While there are obvious correlations between 
the two time intervals in a number of single-pulse flares 
(e.g., event \#53 in Fig.~5a, \#187 in Fig.~5c), in double-pulse flares 
(e.g., event \#367 in Fig.~5i),  or in multi-pulse flares 
(e.g., event \#54 in Fig.~6d, \#150 in Fig.~6e), we see also surprising 
cases where hard X-ray emission
is detected for a single pulse only when a sequence of 5 magnetic
energy pulses is present (e.g., event \#171 in Fig.~6j, \#219 in Fig.~6i).
Thus we find both, well-correlated flare events, as well as 
mismatching time profiles. This outcome of our study indicates
that the simple-minded notion of magnetic energy dissipation with
subsequent particle acceleration does not always fit the data. 

\section{	DISCUSSION					}

\subsection{	Slow-Driven SOC Models 				}

We consider two different scenarios of the time evolution of energy
dissipation in solar flares: the slow-driven {\sl self-organized 
criticality (SOC)} model (Fig.~1a), and the fast-driven SOC model
(Fig.~1b). The slow-driven SOC model corresponds to the model of
cosmic transients proposed by Rosner and Vaiana (1978), while 
their time evolution can 
also be characterized by an exponential-growth model, a power law-growth 
model, or a logistic-growth model (Section 3 of Aschwanden 2011a).
Besides the application to solar flare observations, 
slow external driving of photospheric 
motion is expected to lead to occasional relaxation events also,
at random times, with random amplitudes (Longcope and Sudan 1992).
The essential property of the slow SOC model is the exponential
growth of energy build-up during the time interval between two
subsequent flare events, which eventually creates a flare at a 
random time interval, and relaxes then into a more stable state than
before. The exponential growth function, together with Poissonian
random statistics, leads to the prediction of a power law function 
of the flare size distribution (Rosner and Vaiana 1978). Moreover,
the monotonic growth of the free energy predicts a correlation
between the flare size and the inter-flare (waiting) time interval.
However, observational searches for such a correlation between the
flare sizes and flare waiting times turned out to be negative
(Lu 1995; Crosby et al.~1998; Wheatland 2000a; Moon et al.~2001;
Lippiello et al.~2010). The only correlation found was that 
smaller active regions produce smaller flare sizes (Wheatland 2000b),
and that small active regions produce deviations from power laws (Wheatland 2010).
There are also the problems that large flares sometimes occur within 
shorter waiting times than the required energy build-up times of 
the Rosner-Vaiana model, sometimes a larger flare volume is required 
than available, or too many e-folding growth times are necessary 
(Lu 1995). Nevertheless, a correlation of the flare size with the 
time interval {\sl after} a flare (rather than {\sl before}) was 
claimed for a small sample of flare events in the same active 
region (Hudson 2019). In summary, none of the predictions of the
slow-driven SOC model of Rosner and Vaiana (1978) could be confirmed
by solar flare observations.

\subsection{	Fast-Driven SOC Models			 	}

Most of the simulations of the (frequency occurrence) size distributions
of SOC avalanches assume a separation of time scales, which means that
the avalanche duration $\tau_{flare}$ 
or energy dissipation time scale $\tau_{diss}$ is much shorter 
than the waiting time between two subsequent avalanches, i.e.,
$\tau_{flare} \ll \Delta t_{wait}$. If the input rate 
(e.g., of sand grains dripped on a sand pile) is sufficiently
slow, the statistical properties of avalanche sizes and durations
are expected not to change (Pruessner 2012). However, the observed
statistics of solar flares was found to violate the time scale 
separation during the solar cycle maximum era 
(Aschwanden 2011a, 2011b, 2011c;
Aschwanden and Freeland 2012), when the flare duration exceeded
the waiting times, i.e., $\tau_{flare} \gapprox \Delta t_{wait}$,
which we call a fast-driven SOC system. Since the mean waiting
time $<\Delta t_{wait}>$ is defined by the total duration $T$ of
the observations,  
divided by the total number $N_{ev}$ of events (or intervals),
\begin{equation}
	<\Delta t_{wait}> = {T \over N_{ev}} \ ,
\end{equation}
the mean waiting time decreases reciprocally with the number of
events, and thus becomes shorter for a faster input rate, as
shown in Fig.~3 for a fast driver that has a factor of $10^2$
higher event number, but also a factor of $10^2$ shorter mean
waiting time. As the 10 examples in Fig.~6 demonstrate,
a number of $N_{peak}=5-13$ flare peaks occur in large
flares, which represent elementary flare substructures 
(Aschwanden et al.~1998), that
we interpret as individual energy dissipation events in a 
fast-driven SOC system. Thus, we detect rapid fluctuations 
of the free energy $E_{free}(t)$ before, during, and after 
large flares in a fast-driven SOC system (Figs.~5 and 6), but the
free energy does not 
monotonically increase between two subsequent flares (Fig.~1a).
Hence, the fast-driven SOC model (Fig.~1b) is more consistent 
with the observations than the slow-driven SOC model (Fig.~1a).

\subsection{	The Time Evolution of the Free Energy 	}

If the free (magnetic) energy that is dissipated during a solar
flare would all be stored in the corona, we should see a negatively
dropping step function of the free energy $E_{free}(t)$ during
the flare duration (Fig.~1a). 
One of the most detailed studies on the time evolution of the free
energy shows a gradual build-up of free energy during 2 days, 
culminating with an X2.2 GOES-class flare and a simultaneous
downward step in the free energy (Sun et al.~2012;
Aschwanden et al.~2014b). However, discrepancies up to a factor
of $\lapprox 10$ have been noticed in the decrease of free energy during flares,
when the standard Wiegelmann-NLFFF code (with pre-processing) was
employed in addition to our VCA-NLFFF code (Aschwanden et al.~2014b),
which was reduced down to a factor of $\lapprox 3$ in recent refined 
magnetic modeling (Aschwanden 2020).
Besides the expected step functions, we observe in the present study 
also a number of pulses in the free energy that have a short 
rise time and decay time, in the order of $\tau_{rise} \approx 
\tau_{decay} \approx \tau_{pulse}/2 \approx 0.2\pm 0.1$ hrs = 
12$\pm$6 min (Fig.~8a,b,c).

A puzzling question is what mechanism causes the relatively short 
rise time of the free energy? One mechanism that we know to produce
an increase of the free energy is the helical twisting by vertical
currents (as it is incorporated in the VCA-NLFFF code used here), 
but then the twisting with subsequent un-twisting produces a 
time-symmetric pulse in the free energy without net energy transfer.
Another possible mechanism is the {\sl coronal illumination effect},
where the twisted loops are not visible in the initial flare phase,
but become detectable when chromospheric evaporation starts to fill
up the flare loops (Aschwanden et al.~2014a). A third possibility is
chromospheric energy injection into the corona produced by energy
transferred from the turbulent convection zone and photosphere 
into the corona, e.g., via anomalous current dissipation (Rosner et al.~1978). 
Such a scenario with the ultimate energy source in the convection zone 
rather than in the corona, can draw large amounts of free energy for 
generating a flare without requiring coronal storage. 
Magneto-convection as seen in photospheric granulation cells has 
typical spatial scales of $\approx 1000$ km and turnover times of
$\approx 7$ min, which produces new emerging flux on time scales
close to the observed pulse rise times of $\tau_{pulse} \approx 12\pm6$ min
(Fig.~8a). In conclusion, the time evolution of the free energy $E_{free}(t)$
provides crucial constraints how and where the flare energy is stored.

\subsection{	Non-Stationary Driver and Waiting Time		}						 
From the waiting time distribution we can learn whether a SOC system
is stationary or non-stationary, which means whether the mean flaring
rate is constant or not, as a function of time. In the original
SOC concepts of Bak et al.~(1987) it was assumed that individual 
avalanches are statistically independent events, and thus the
waiting time distribution should form a Poissonian (or exponential)
distribution function. If there is a deviation from a Poissonian
distribution apparent, individual avalanche events could not be
independent events, such as in sympathetic flares 
(Wheatland 2002, 2006; Wheatland and Craig 2006; Moon et al. 2002, 2003).
However, when the flaring rate is not constant,
the resulting waiting time distribution can be calculated by 
summing the partial waiting time distributions for each flaring rate
(Wheatland et al.~1998; Wheatland and Glukhov 1998; Wheatland 2000c)
as we summarize in Section 2.1 (and in Section 5 of Aschwanden 2011a). 
Waiting time distributions of solar flare data generally show a
power law distribution with a slope of $\alpha_{\Delta w} \approx 2-3 $,
(Wheatland et al.~1998; Wheatland 2003; Moon et al.~2001; 
Kanazir and Wheatland 2010; Aschwanden and McTiernan 2010),
which is explained here with a model that is based on
on the universal reciprocal relationship between the (time-varying) 
mean flaring rate and the (time-varying) waiting time, and predicts 
a slope of $\alpha_{\Delta t}=2$. 
In summary, the non-stationary Poissonian model provides the most 
natural explanation for the observed power law-like
waiting time distributions. 
 
Besides the non-stationary Poissonian model of a fast-driven SOC model, 
some alternative interpretations were explored too. An energy balance 
model in terms of a {\sl master equation} between energy build-up and
energy loss by dissipation of free energy has been proposed 
(Wheatland and Glukhov 1998; Wheatland and Litvinenko 2001, 2002;
Wheatland 2008; 2009; Wheatland 2009),
Other approaches use scaling laws from magnetic reconnection processes 
(Wheatland and Craig 2003, 2006).
Alternative functions for waiting time distributions were tested also,
finding that lognormal and inverse gaussian distribution functions are
more likely to fit the observations than the exponential function (Kubo 2008).  

\section{		CONCLUSIONS 					}

Standard {\sl self-organized criticality (SOC)} models, mostly inspired
by the paradigm of sandpile avalanches introduced by Bak et al.~(1987), 
assume a slow-driven energy dissipation system, a stationary energy input 
rate, a fixed (critical) threshold for triggering of avalanches, 
time scale separation between avalanche time durations $\tau_{dur}$
and inter-event waiting times $\Delta t$, i.e., $\tau_{dur} \gg \Delta t$,
and statistical independence of individual avalanche events. These
assumptions predict power law distribution functions for most avalanche 
parameters (such as the size and duration) and exponential distributions 
for the waiting times. In reality, however, most of these assumptions are 
violated, but it appears that SOC models are sufficiently robust to 
preserve some power law characteristics, even in the presence of violated 
assumptions. In this study we explore non-standard SOC models that 
account for the violated assumptions, in particular for the 
phenomenon of solar flares. Our findings are the following:

\begin{enumerate}
\item{\underbar{The waiting time distribution:}
One not understood problem is the functional shape of the waiting
time distribution, because the assumption of statistical independence of
individual avalanche events predicts an exponential function, while the
observations exhibit a power law distribution with a slope of $\alpha_{\Delta t}
\approx 2-3$. One possible solution of this problem is the {\sl
non-stationary Poisson} model, introduced by Wheatland and Litvinenko (2002),
but the functional shape of the flaring rate $\lambda(t)$ has not been
constrained. The shape of observed waiting time distributions has been 
reconciled empirically with the near-reciprocal flaring rate function
$f(\lambda) = \lambda^{-1} \exp{(-\lambda/\lambda_0)}$ (Eq.~6) in the previous 
study of Aschwanden and McTiernan (2010). In the present study we provide 
a physical reason in terms of the universally valid reciprocal relationship 
between the mean flaring rate $<\lambda>$ and the mean waiting time
$<\Delta t>$, i.e., $f(\lambda) \propto <\lambda>^{-1} = <\Delta t>$.
The reciprocal relationship predicts then a power law distribution for 
the waiting time distribution, with a power law slope of 
$\alpha_{\Delta t}=2$, without any free parameters, except for a 
normalization constant $\lambda_0$.}

\item{\underbar{Non-stationarity of SOC model:}
The power law shape of the waiting time distribution thus yields a sufficient 
(but not necessary) condition for the non-stationarity of the flare rate 
that drives the generation of solar flares. The flare rate varies up to
two orders of magnitude between the minimum or maximum of the solar 
magnetic (Hale) cycle. There are also large variations in the flaring rate
on shorter time scales, down to weeks, days, or hours.
All this variability produces power law-like distributions
of waiting times. Moreover, it produces also power law distributions 
for the size and durations of flares, which appears to be a very robust 
feature of SOC models, regardless whether the driver is stationary or 
non-stationary.} 

\item{\underbar{Slow-driven and fast-driven SOC models:} While the 
duration of an avalanche (e.g., a solar flare) is much shorter than 
the waiting time between two subsequent avalanche events in standard 
SOC models, we find that this behavior is only true in quiescent periods
during the solar cycle minimum, especially when the SOC threshold
is high and the flaring rate is low. However, the flare rate during
solar maximum conditions is often so high that near-simultaneous
flare events overlap in time and thus the flare duration becomes
comparable with the waiting time or even exceeds the waiting time. 
The solar dynamo thus produces a SOC system that oscillates between
slow-driven and fast-driven operation cycles.}

\item{\underbar{The Rosner-Vaiana (1978) model:} This model predicts
a continuously growing energy storage between two flare events, and
thus a correlation between the waiting time and dissipated energy
during the following event. Observations do not confirm 
that energy is stored between two flares, nor is there any 
correlation between storage time and energy dissipation. Although
we can measure free (magnetic) energy before, during, and after
flares, we rarely see a simple step function of the free energy 
that drops from a high pre-flare level to a low post-flare level.}

\item{\underbar{Pulsed free energy dissipation:} Instead of a
step function in the time evolution of the free energy, we observe
that the free energy exhibits pulses with rise times and decay
times of $\approx 12\pm6$ min, which occur between 1 and 13 times
during a flare, depending on the flare duration (1.1-5.2 hrs).
The fact that each pulse exhibits a fast rise (rather than a slow
rise as expected in storage models), indicates that free energy
is intermittently generated (rather than stored over long time
intervals), for instance by photospheric convection, which shows
similar turnover times of order $\approx 7$ min in the photospheric
granulation layer.} 

\end{enumerate}

Based on these results we recommend to modify numerical simulations 
of SOC models with the following features, in order to obtain a more
realistic representation of solar flare data: (i) A non-stationary
driver that varies from slow-driven dynamics during the solar minimum,
to fast-driven dynamics during the solar maximum; (ii) Separate
fitting of time periods with low and high flaring rates, possibly
measuring the flaring rate distribution $\lambda(t)$ as a function
of time; (iii) Fitting of the predicted waiting time distribution
model $p(\Delta t)=\lambda_0/(1+\lambda_0 \Delta t)^2$ (rather than
fitting a straight power law function); (iv) Localization of
photospheric convection vortices during flares in magnetogram data 
that contribute most significantly to local increases in the
free energy during flares; and (v) spatio-temporal disentangling
of near-simultaneous flare sites during fast-driven time periods.

\acknowledgements
Part of the work was supported by
NASA contract NNG 04EA00C of the SDO/AIA instrument and
the NASA STEREO mission under NRL contract N00173-02-C-2035.

\section*{REFERENCES} 

\def\ref#1{\par\noindent\hangindent1cm {#1}}

\ref{Aschwanden, M.J., Dennis, B.R., and Benz, A.O.
	1998, ApJ 497, 972.}
\ref{Aschwanden, M.J. and McTiernan, J.M.
 	2010, ApJ 717, 683.}
\ref{Aschwanden, M.J.
 	2011a, {\sl Self-Organized Criticality in Astrophysics. 
	The Statistics of Nonlinear Processes in the Universe},
 	Springer-Praxis: New York, 416p.}
\ref{Aschwanden, M.J.
	2011b, SoPh 274, 99.}
\ref{Aschwanden, M.J.
	2011c, SoPh 274, 119.}
\ref{Aschwanden, M.J. and Freeland, S.L.
 	2012, ApJ 754, 112.}
\ref{Aschwanden, M.J. 
	2013, SoPh 287, 323.} 
\ref{Aschwanden, M.J., Xu, Y., and Jing, J. 
	2014a, ApJ 797:50.} 
\ref{Aschwanden, M.J., Xu, Y., and Y. Liu 
	2014b, ApJ 785:34.}
\ref{Aschwanden, M.J., Crosby, N., Dimitropoulou, M., Georgoulis, M.K., 
	Hergarten, S., McAteer, J., Milovanov, A., Mineshige, S., 
	Morales, L., Nishizuka, N., Pruessner, G., Sanchez, R., Sharma, S., 
	Strugarek, A., and Uritsky, V.
 	2016a, SSRv 198, 47.}
\ref{Aschwanden et al. 
	2016b, ApJSS 224, 25.}
\ref{Aschwanden, M.J. 
	2019, {\sl New Millennium Solar Physics}, Springer Nature, 
	Switzerland, Science Library Vol. 458.}
\ref{Aschwanden, M.J. 
	2020, ApJ (subm.)} 
\ref{Bak, P., Tang, C., and Wiesenfeld, K.
 	1987, PhRvL 59/4, 381.}
\ref{Charbonneau, P., McIntosh, S.W., Liu,H.L., and Bogdan,T.J.
 	2001, SoPh 203, 321.}
\ref{Charbonneau, P.
 	2013, {\sl Self-Organized Criticality and Solar Flares},
 	Chapter 12 in {\sl Self-Organized Criticality Systems},
	(ed. Aschwanden,M.J.), Open Academic Press GmbH \& Co., 
	http://www.openacademicpress.de, p.371.}
\ref{Crosby, N.B., Aschwanden, M.J. and Dennis, B.R.
 	1993, SoPh 143, 275.}
\ref{Crosby, N., Vilmer, N., Lund, N., Sunjaev, R. 
	1998, A\&A 334, 299.}
\ref{Dennis, B.R. and Zarro, D.M.
	1993, SoPh 146, 177.}
\ref{Hudson, H.S. 
	2019, MNRAS, (subm).}
\ref{Kanazir, M. and Wheatland, M.S.
	2010, SoPh 266, 301.}
\ref{Kubo, Y.
	2008, SoPh 248, 85.}
\ref{Lippiello, E., de Arcangelis, L., and Godano, C.
 	2010, A\&A 511, L2.}
\ref{Litvinenko, Y.E.
 	1996, SoPh 167, 321.}
\ref{Litvinenko, Y.E.
	1996, SoPh 167, 321.}
\ref{Litvinenko, Y.E. and Wheatland, M.S.
 	2001, ApJ 550, L109.}
\ref{Longcope, D.W. and Sudan, R.N.
	1992, Phys.Fluids B: Plasma Physics 4/7, 2277.}
\ref{Lu, E.T. and Hamilton, R.J.
 	1991, ApJ 380, L89.}
\ref{Lu, E.T.
	1995, ApJ 447, 416.}
\ref{Jaynes, E.T.
	2003, {\sl Probability Theory: The Logic of Science},
	Cambridge University Press: Cambridge, 758p.}
\ref{Moon, Y.J., Choe, G.S., Yun, H.S., and Park, Y.D.
	2001, JGR 106/A12, 29951.}
\ref{Moon, Y.J., Choe, G.S., Park, Y.D., Wang, H., Gallagher, P.T., 
	Chae, J., Yun, H.S. and Goode, P.R.
 	2002, ApJ 574, 434.}
\ref{Moon, Y.J., Choe, G.S., Wang, H., and Park,Y.D.
 	2003, ApJ 588, L1176.}
\ref{Neupert, W.M.
	1968, ApJ 153, L59.}
\ref{Pesnell, W.D., Thompson, B.J., and Chamberlin, P.C.
 	2011, SoPh 275, 3.}
\ref{Pruessner, G.
 	2012, {\sl Self-organised criticality. Theory, models 
	and characterisation}, ISBN 9780521853354, Cambridge 
	University Press, Cambridge}
\ref{Rosner, R. and Vaiana, G.S. 
	1978, ApJ 222, 1104.}
\ref{Rosner, R., Golub, L., Coppi, B., and Vaiana, G.S.
 	1978, ApJ 220, 643.}
\ref{Scargle, J.
	1998, ApJ 504, 405.}
\ref{Sivia, D.S. and Skilling, J.
	2006, (2nd ed.), {\sl Data Analysis - A Bayesian Tutorial},
	Oxford University Press, 264p.}
\ref{Sun, X., Hoeksema, J.T., Liu, Y., Wiegelmann, T., Hayashi, K., 
	Chen, Q., and Thalmann, J.
 	2012, ApJ 748, 77.}
\ref{Wheatland, M.S. and Glukhov, S.
	1998, ApJ 494, 858.}
\ref{Wheatland, M.S., Sturrock, P.A., and McTiernan, J.M. 
	1998, ApJ 509, 448}
\ref{Wheatland, M.S. 
	2000a, ApJ 532, 1209.}
\ref{Wheatland, M.S. 
	2000b, SoPh 191, 381.}
\ref{Wheatland, M.S. 
	2000c, ApJ 536, L109.}
\ref{Wheatland, M.S.  Sturrock, P.A. and Roumeliotis, G. 
	2000, ApJ, 540, 1150.}
\ref{Wheatland, M.S. and Litvinenko, Y.E.
	2001, ApJ 557, 332.}
\ref{Wheatland, M.S. 
	2002, SoPh 208, 33.}
\ref{Wheatland, M.S. and Litvinenko, Y.E.
	2002, SoPh 211, 255.}
\ref{Wheatland, M.S. and Craig, I.J.D.
	2003, ApJ 595, 458.}
\ref{Wheatland, M.S.
	2003, SoPh 214, 361.}
\ref{Wheatland, M.S. 
	2006, SoPh 236, 313.}
\ref{Wheatland, M.S. and Craig, I.J.D.
	2006, SoPh 238, 73.}
\ref{Wheatland, M.S.
	2008, ApJ 679, 1621.}
\ref{Wheatland, M.S.
	2009, SoPh 255, 211.}
\ref{Wheatland, M.S.
	2010, ApJ 710, 1324.}

\clearpage


\begin{figure}
\centerline{\includegraphics[width=0.8\textwidth]{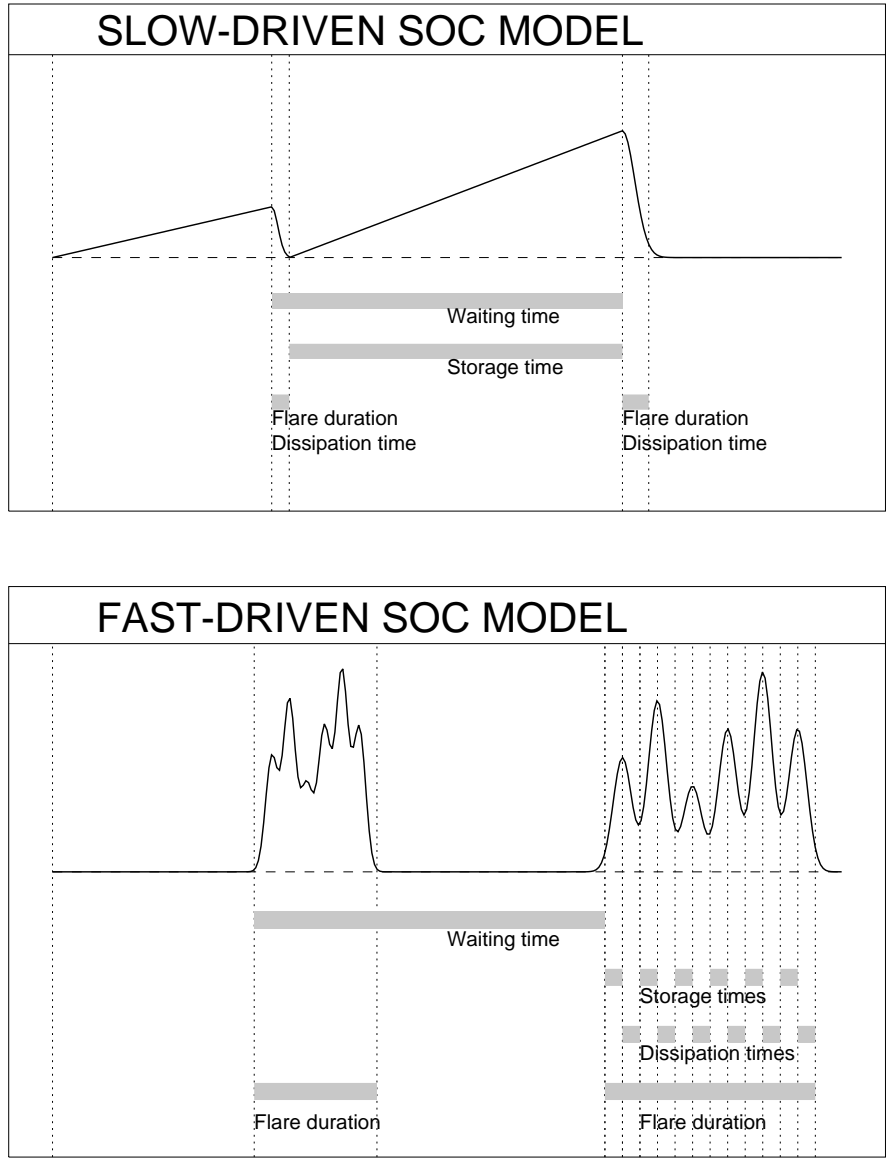}}
\caption{Definition of time scales for the slow-driven SOC model,
according to the Rosner and Vaiana (1978) model
(top panel), and the fast-driven SOC model proposed in this study
(bottom panel).
The x-axis represents the time, and the y-axis represents 
the time evolution of the free energy $E_{free}(t)$ that is
dissipated during flares.}
\end{figure}

\begin{figure}
\centerline{\includegraphics[width=1.0\textwidth]{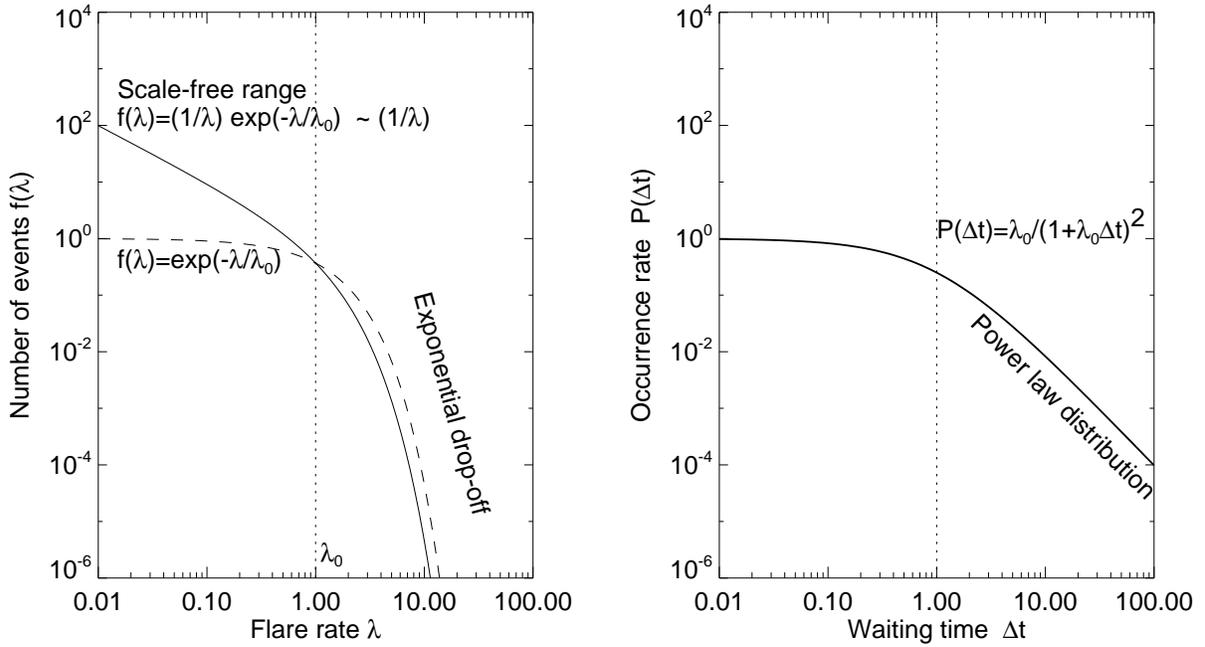}}
\caption{{\sl Left:} The distribution function $f(\lambda)$ 
of the flaring rate is approximately reciprocal in the scale-free
range $\lambda \lapprox \lambda_0$ and shows a steep exponential
drop-off at larger flaring rates $(\lambda \gapprox  \lambda_0)$.
{\sl Right:} The resulting waiting time distribution is predicted
to have a power law function $P(\Delta t) \propto \Delta t^{-2}$
at large waiting times $\Delta t \gapprox \Delta t_0$, shown 
here for a value of $\lambda_0=1$.} 
\end{figure}

\begin{figure}
\centerline{\includegraphics[width=1.0\textwidth]{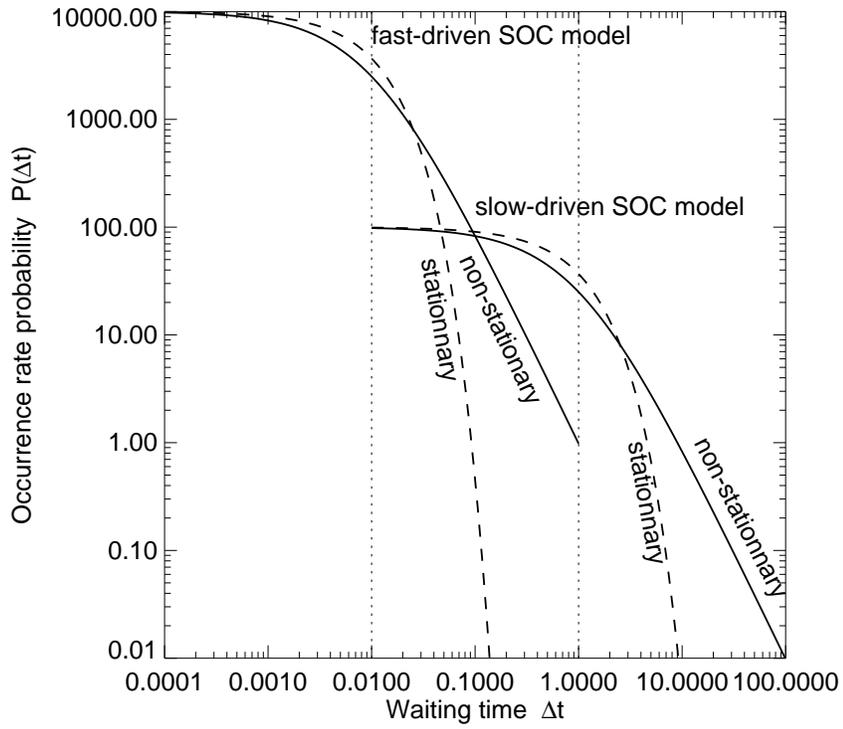}}
\caption{Overview of waiting time distributions for slow-driven 
and fast-driven, stationary and non-stationary SOC models.}
\end{figure}

\begin{figure}
\centerline{\includegraphics[width=1.0\textwidth]{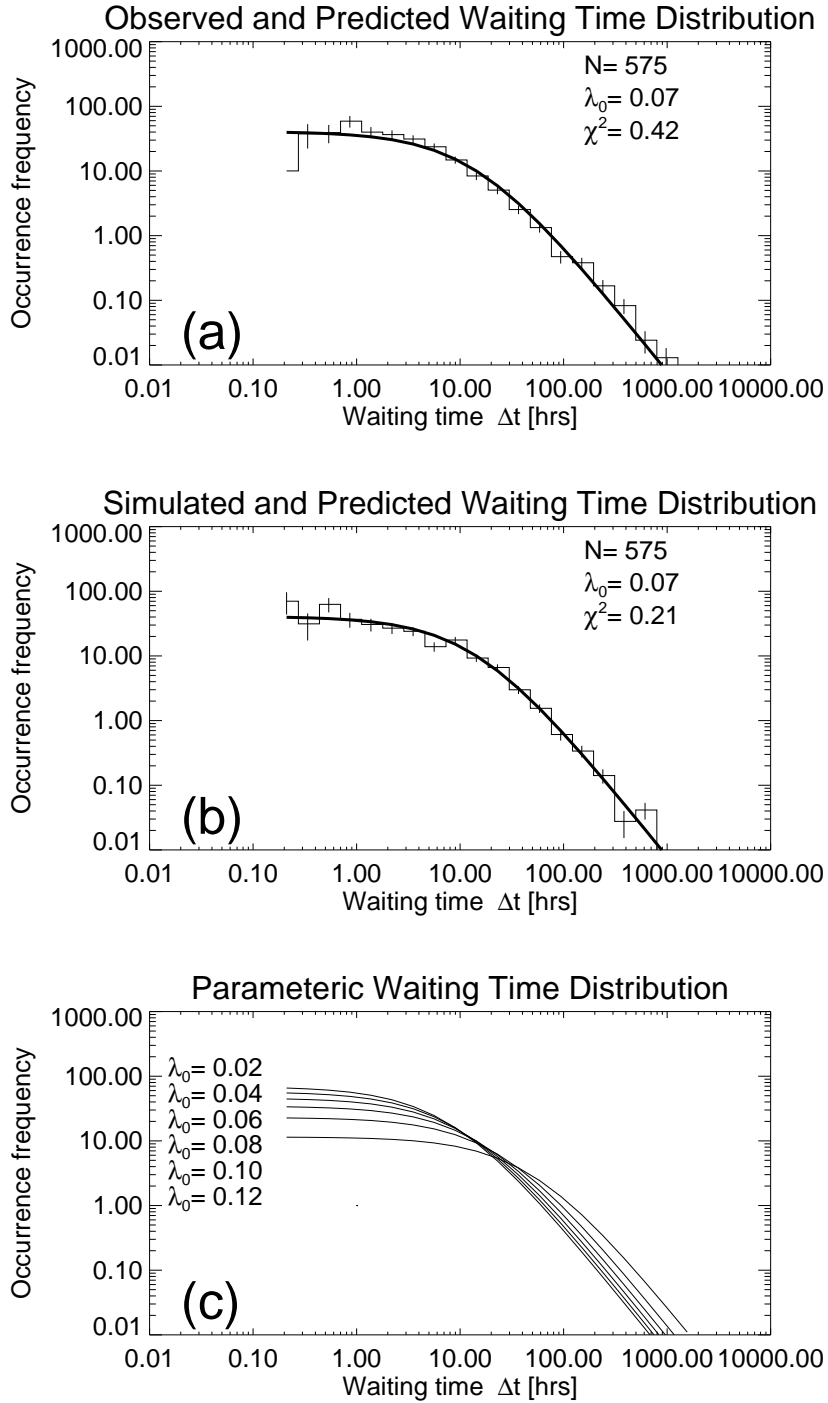}}
\caption{(a) Observed waiting time distribution of GOES
M- and X-class flares (histogram) with predicted model (thick
solid curve); (b) Simulated waiting time distribution
for the same normalization constant $\lambda_0=0.07$ (histogram)
with predicted model (thick solid curve); and (c)
Parametric set of waiting time distributions for
$\lambda_0=0.02, 0.04, ..., 0.12$ hrs$^{-1}$.}
\end{figure}

\begin{figure}
\centerline{\includegraphics[width=1.0\textwidth]{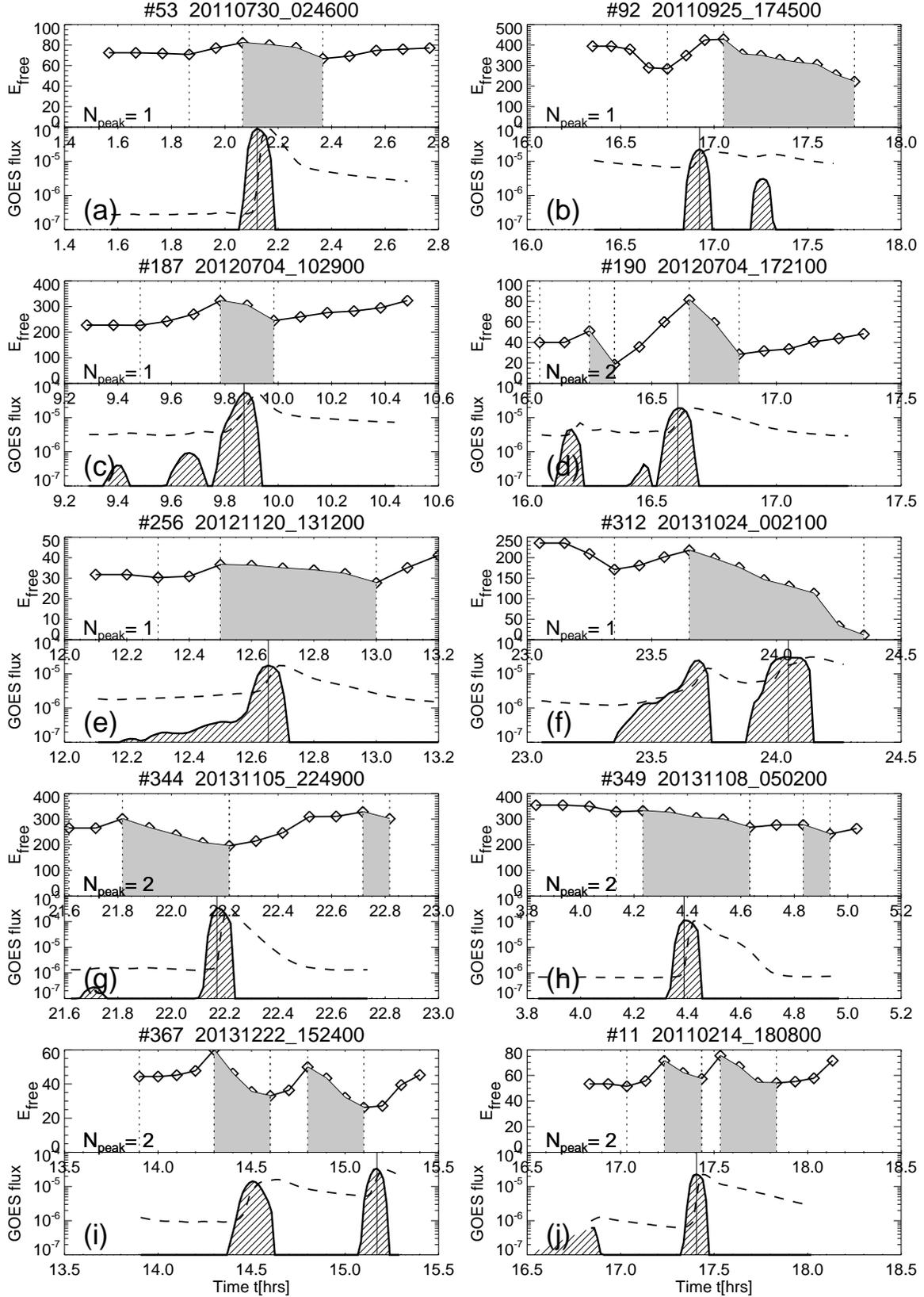}}
\caption{The time evolution of the free energy $E_{free}(t)$
in 10 flares with one or two peaks of the energy energy loading/dissipation 
episodes (thick black curves with diamonds). The GOES flux curve
is indicated with a dashed curve, and the time derivative of the
GOES curve with a solid line with hatched areas. 
The time intervals of energy dissipation are colored in grey.}
\end{figure}

\begin{figure}
\centerline{\includegraphics[width=1.0\textwidth]{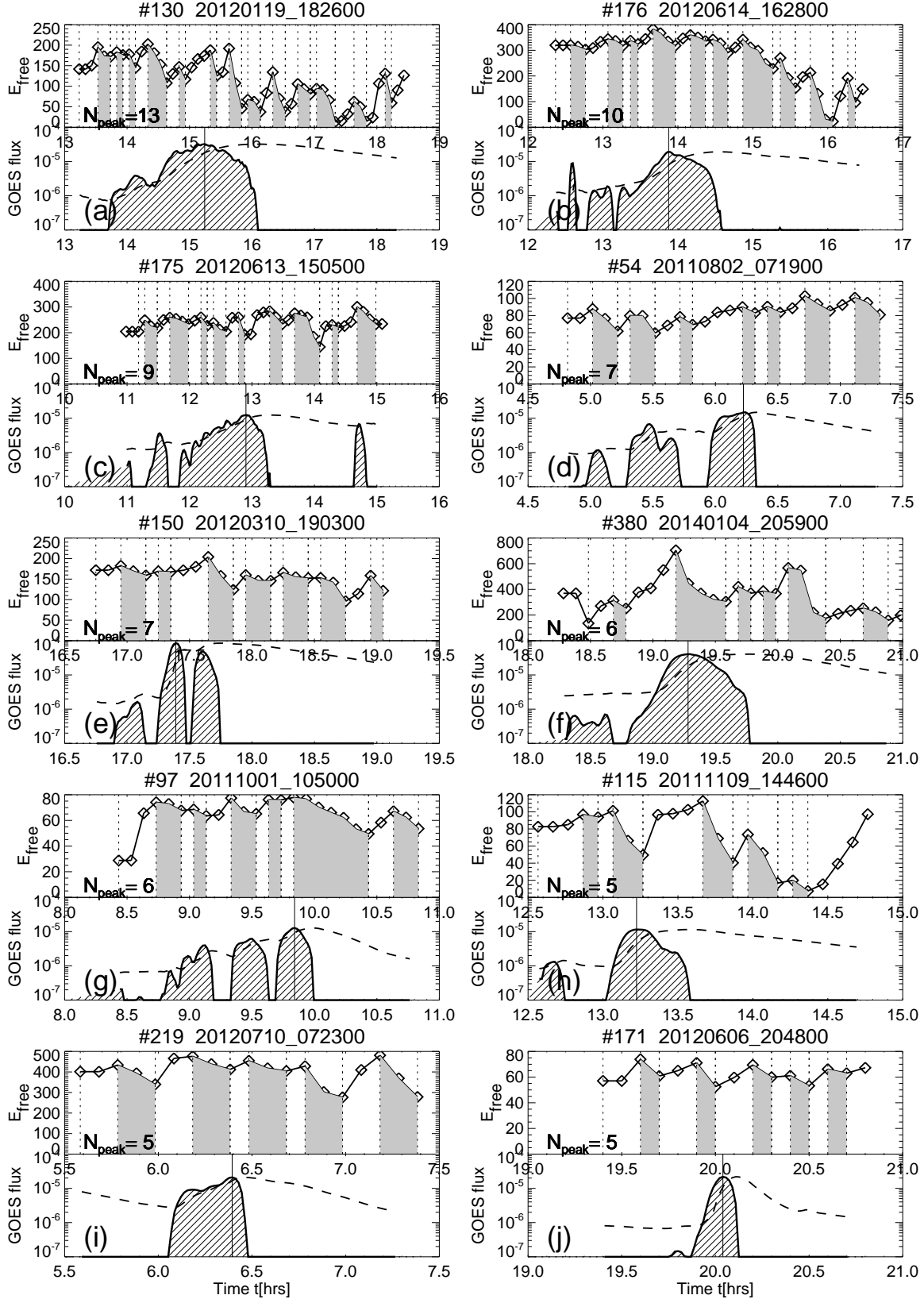}}
\caption{The time evolution of the free energy $E_{free}(t)$
in 10 flares with the largest number of energy loading/dissipation 
episodes ($N_{peak}=5-13$) (thick black curves with diamonds).  
The time intervals of energy dissipation are colored in grey.
Otherwise similar presentation as in Fig.~5.}
\end{figure}

\begin{figure}
\centerline{\includegraphics[width=1.0\textwidth]{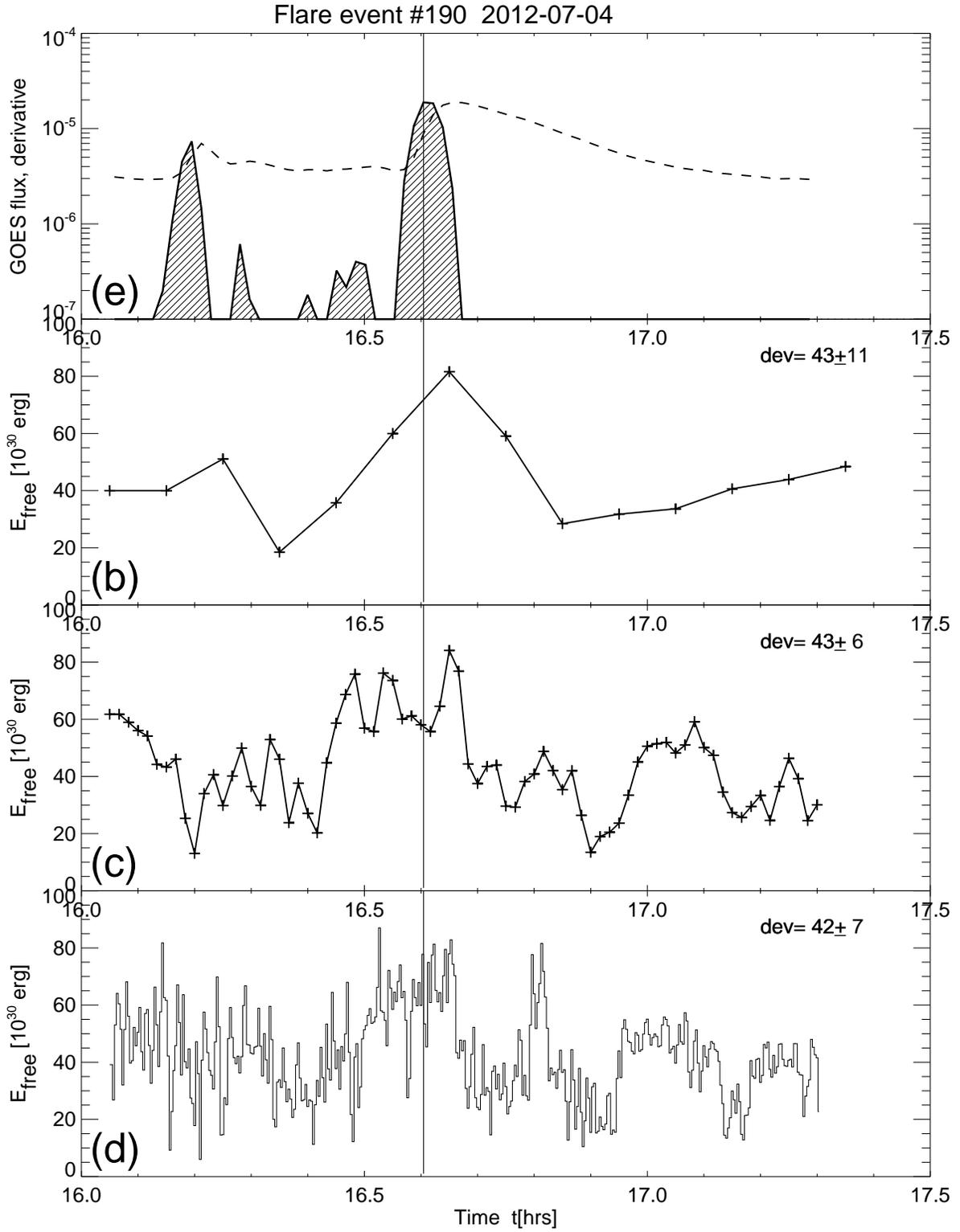}}
\caption{(a) GOES flux (dashed curve) and time derivative
(hatched curve); The evolution of the free energy is shown
with different time resolutions: (b) 6-minute cadence;
(c) 1-minute cadence; (d) 12-s time cadence.} 
\end{figure}

\begin{figure}
\centerline{\includegraphics[width=1.0\textwidth]{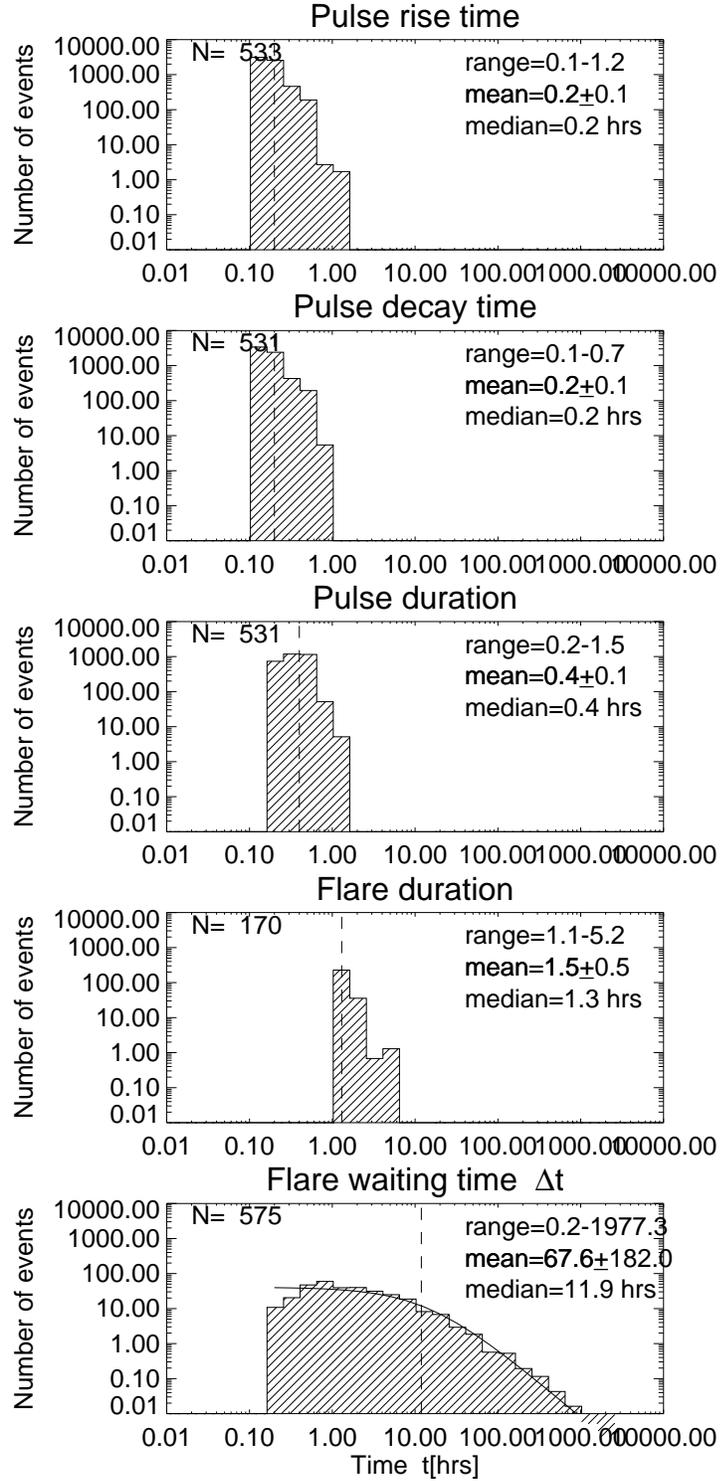}}
\caption{Number of events (per bin) as a function of
(a) the free energy pulse rise time, (b) pulse decay time,
(c) pulse durations, (d) flare duration, and flare waiting time
distribution (bottom panel), along with the theoretical model
for $\lambda_0=0.07$ (curve).}
\end{figure}

\end{document}